\documentclass[seceq]{ptptex}

\newcommand{\vect}[1]{\mbox{\boldmath$#1$}}

\renewcommand*{\theequation}{\thesection.\arabic{equation}}
\usepackage{graphicx}
\usepackage{calligra}
\usepackage{calrsfs}
\usepackage{mathrsfs}
\title{General Relativistic Hartree-Fock Calculations for Neutron Star}

\author{Naoki \textsc{Onishi},
}
\inst{
Department of Physics, Tokyo Metropolitan University,\\ Minamiohsawa, Hachioji
192-0397, Japan 
}

\abst{
We investigate the global structure of neutron stars within the framework of general relativity,
treating the entire star as a quantum-degenerate system.
Rather than relying on the Tolman-Oppenheimer-Volkoff (TOV) equation,
we solve the Einstein-Cartan (EC) field equations self-consistently, incorporating  
the energy-momentum tensor contributions from neutrons.
Neutron wave functions are obtained by solving the Dirac equation in a curved spacetime
with both torsion and curvature effects.
Given that neutron star stars contain about 10$^{57}$ particles, we adopt a 
scaled $\hbar$ approach to efficiently describe the quantum state of this highly degenerate system. 
}

\begin{document}
\maketitle
 
\section{Introduction}\label{intro}
 Theoretical msodels of neutron stars (NSs) have traditionally employed TOV) formalisms, treating
the star either as a   classical fluid \cite{To39} or Fermi gas \cite{OV39} (see \S \ref{NmCl}).
In the TOV framework, the energy-momentum tensor is locally diagonal
 and enters Einstein's equation through the equation of state (EOS).
 However, this approach neglects quantum features, such as phase relations
 between the states in the adjacent boxes of quantization.
 
 In this study we approach NSs as quantum-degenerate systems where long range correlations
 arise from the anti-symmetrization of fermion wave functions.
 The star's interior contains neutrons, protons and electrons in $\beta$-stable equilibrium, governed by
 the chemical potential condition:
 $\mu_{\rm n}=\mu_{\rm p}+\mu_{\rm e}$ where $\mu_{\rm n}$ ,$\mu_{\rm p}$ and $\mu_{\rm e}$ 
 represent the chemical potentials of neutrons, protons and electrons, respectively.
 When $\mu_{\rm n}>\mu_{\rm p}+\mu_{\rm e}$, high energy neutrons decay via
${\rm n} \rightarrow {\rm p} + {\rm e}^{-}\hspace{-0.1cm}+\overline{\nu}$, leaving the electron
 above the top of their Fermi sea.
 Conversely, when $\mu_{\rm n}<\mu_{\rm p}+\mu_{\rm e}$, high energy electrons are captured
 by protons through the process
 ${\rm p} + {\rm e}^{-} \hspace{-0.1cm} \rightarrow {\rm n} +\nu$,
 placing neutrons at the top of the neutron's Fermi sea.
 These processes maintain $\beta$-equilibrium stabilized by Pauli-blocking effects from the quantum
 degeneracy. 
 
 To account for these quantum effects, we employ the Hartree-Fock(HF) method. In this approach,
 the total wave function is constructed from three Slater determinants corresponding to
 the wave functions of neutrons, protons and electrons, respectively.  
 For illustrative purposes, a typical size of NS having mass and radius,
 1.5 $M_{\odot}$ and $R=12$ Km, respectively.
 The number of nucleons, mostly neutrons, is about 1700$\times N_{\star}$,
 where $N_{\star}$ stands for a big number of 
 $10^{54}  (=10^{3\times 18}$ coming from 1km$= 10^{18}$ fm).
The average nucleon density turns out to be 0.25$N_{\star}$km$^{-3}$=0.25 fm$^{-3}$,
greater than that in nucleus (0.17 fm$^{-3}$), and therefore the Fermi energy of neutron is about 75 MeV.
It is, of course, unfeasible to treat numerically such a gigantic number of particles in microscopic ways 
and therefore it is employed the scaled hbar method (S$\hbar$M) changing the unit of action
by scaling factor $\gamma$, like $\hbar_{*} =\gamma \hbar_{\star}$ with $\hbar_{\star}=10^{18}\hbar$. 
One should distinguish $*$ from $\star$. 

This S$\hbar$M is based on the following notion.
In a single-Slater-determinant, the number of particles and the number of quantum states
are necessarily equal due to Pauli principle.
Therefore, if the scaled Planck constant $\hbar_{*}$ is used in Dirac equation, 
the wave function so obtained describes the wave function of assemblage
constituted by $\gamma^{3} N_{\star}$ Fermions.
The resultant physical variables such as density, current density, etc., become the more accurate
by using the smaller scaling factor $\gamma$,
because the values of quantum numbers,
such as wave number like radial node number and angular momenta,  
become the larger inversely the smaller scaling factor, resulting total wave function well described due to
the correspondence principle valid in enough large integer of quantum number. 
In a different light, Dirac field in a quantum degenerate system
may be considered such as classical field.

We also realized that the assemblage corresponds to a kind of mode of collective motion
such as rotational motion.
From the above considerations, it is suitable to choose the scaling factor $\gamma$
in considering size to collectivity \cite{OM20}. 
Therefore, one should take a smaller scaling factor $\gamma$ than the value of
the factor determining the size of collectivity.
Actually, we have examined the accuracy of the calculation by changing the scaling factor
and comparing the results obtained using different scaling factors $\gamma$ \,\cite{OM20}.
In this paper as the first trial, we focus on only neutron system in the gravitational field,
without taking into account electromagnetic and nuclear interactions,
which bring about other complexities, even though play important roles
in constructing the characteristic structure of in NS.
These contents may be discussed in coming papers including magnetic structure\cite{OM20} . 

In section \ref{DECE}, we explain the procedure of iterative three steps calculations, namely
(1) solve Dirac equation in gravitation field with S$\hbar$M,
(2) calculate energy-momentum tensor and spin current using the wave functions obtained 
in the first step,
(3) solve Einstein-Cartan field equation (EC) including the first and second structure equations of
torsion and curvature to calculate gravitational potentials used in the first step
of the next cycle of the iteration. 
We treat torsion because neutron, proton and electron are spinors, which bring about naturally
the torsion through spin-current. 

In section \ref{Rp_vsh}, in taking the fact into account that NS is likely to keep 
spherical due to the strong gravitational force, the three equations, i.e., Dirac, energy-momentum,
and EC, are reformulated  in terms of spinor spherical harmonics with the aid of Racah algebra.
In section \ref{NmCl} the numerical calculations of the iteration of cyclic procedure is demonstrated
how to converge the wave functions and the gravitational potentials, in the assumption of spherical
symmetry.
The resultant density distributions are compared to those given by the TOV calculations.
In Sec.\ref{dscsn}  we will sketch briefly a picture related to global structure of NS including
protons and electrons with taking account of electromagnetic field.  
%   -------------
%    section 2
%   -------------
\section{Dirac and Einstein-Cartan Equations}
\label{DECE}
In this section, we employ notations of Clifford valued differential form just
for a simple description of equations, especially without being bothered
by tedious suffixes and superscfripts. 
 These representations are somewhat abstract, so that, in the next section
 we shall describe them in more practical
 forms adequate for numerical calculations, by using the technique of Racah algebra
 which is necessarily accompanied by complexity.

Clifford numbers were introduced into physics  to obtain square root of 
the Klein-Gordon operator as the following Dirac's $\gamma$-matrices\cite{Di28}
\begin{equation}
\left[ \gamma^{\alpha},\gamma^{\beta}\right]_{+} \equiv
\gamma^{\alpha}\gamma^{ \beta} +\gamma^{\beta}\gamma^{ \alpha}=2\eta^{\alpha \beta}\vect{1}_{4}
\, ,
\hspace{1.0cm} \alpha,\beta=0,1,2,3 \, ,
\end{equation}
where $\vect{1}_{4}$ stands for 4-dimensional unit matrix and 
Minkowski metric signatures are given by
$\eta^{\alpha \beta}\!\!\!=\!\!\eta_{\alpha \beta}\!\!=\!\!{\rm diag} (1,-1,-1,-1)$
acting to lower and raise indices, like
$\gamma_{\alpha}\!\!=\!\eta_{\alpha \beta}\gamma^{\beta}$ .
Four elements of the vector $\gamma^{\alpha} $ (hereafter called D-vector) generate the other 11 linearly independent Clifford numbers, 
namely, 6 elements of antisymmetric tensor (D-tensor)
$\sigma^{\alpha \beta}\!\!=\!i \gamma^{\alpha} \gamma^{\beta}=
\eta^{\alpha \gamma}\sigma_{\gamma}^{\,\,\,\beta}$, pseudo scalar 
$\gamma^{5}\!\!=\! i\gamma^{0}\gamma^{1}\gamma^{2}\gamma^{3}$
($\gamma_{5}\!\equiv\!\gamma^{5}\!=\!-i\gamma_{0}\gamma_{1}\gamma_{2}\gamma_{3}$
 as a definition) and 4 elements of the pseudo vector which are written as
\begin{equation}
\begin{array}{l}
 \hspace{0.7cm}  \gamma^{5}\gamma^{\alpha}
=-i\delta^{\alpha}_{\,\,\,\beta \gamma \delta}\gamma^{\beta}\gamma^{\gamma}\gamma^{\delta} ,
\hspace{1.0cm} {\rm with}  \vspace{0.2cm} \\
\delta^{0}_{\,\,\,123}=\delta^{1}_{\,\,\,023}=\delta^{2}_{\,\,\,031}=\delta^{3}_{\,\,\,012}=1
\hspace{0.3cm} {\rm otherwise} \hspace{0.2cm}\delta^{\alpha}_{\,\,\,\beta \gamma \delta}=0 .
\end{array}
\end{equation}
The 4-elements of vector $\gamma^{\alpha}$ and the 6-elements of tensor $\sigma^{\alpha \beta}$
 are represented in terms of two dimensional unit $\vect{1}$ , zero $\vect{0}$
 and Pauli spin  $\sigma_{a}$ matrices as\cite{He00} for D-vector
\begin{equation}
\gamma^{0} = \beta \equiv 
\left( \begin{array}{cc}  \vect{1}  & \vect{0} \\ \vect{0} & -\vect{1} \end{array} \right) ,
\hspace{0.3cm}
\gamma^{a}=\beta \alpha_{a}, \hspace{0.8cm} {\rm with} \hspace{0.2cm}
\alpha_{a}\equiv
\left( \begin{array}{cc}  \vect{0} & \sigma_{a}  \\  \sigma_{a}  & \vect{0} \end{array} \right) ,
\end{equation}
and for D-tensor 
\begin{equation}
\sigma^{ab} =
\left( \begin{array}{cc}  \sigma_{c}  & \vect{0} \\\vect{0} & \sigma_{c} \end{array} \right) \equiv \Sigma^{c}
, \hspace{0.6cm}
 a,b,c=1,2,3 \,\,\,\, {\rm in\,\,cyclic \,\,order} ,
 \end{equation}
 and
 \begin{equation}
\sigma^{0a}=i\left(
\begin{array}{cc} \vect{0} & \sigma_{a} \\ \sigma_{a} & \vect{0} \end{array} \right)=i\gamma^{5}\Sigma^{a}
\equiv\, ^{*}\!\Sigma^{a}
\hspace{0.6cm} {\rm with}  \hspace{0.10cm} \,
\gamma^{5}=
\left( \begin{array}{cc} \vect{0} & \vect{1} \\ \vect{1} & \vect{0} \end{array} \right) \, .
\end{equation}
These six elements of the D-tensor, known as Lorentz rotation 
$\Sigma^{a}$ and boost $^{*}\Sigma^{a}$, constitute Lie algebra like
\begin{equation}
\begin{array}{r}
\left[ \Sigma^{a},\Sigma^{b} \right]_{-}= 2i\Sigma^{c}\, , \hspace{0.2cm}
\left[ \Sigma^{a},  ^{*}\!\!\Sigma^{b} \right]_{-}= 2i \, ^{*}\!\Sigma^{c}\, , \hspace{0.2cm}
 \left[  ^{*}\!\Sigma^{a}, ^{*}\!\!\Sigma^{b} \right]_{-}= -2i \Sigma^{c} ,
\vspace{0.3cm} \\
\hspace{0.3cm} a,b,c=1,2,3 \,\, {\rm in\,\,cyclic \,\,order} \, .
\end{array}
\end{equation}
Hence the theory of gravity can be treated as a gauge theory\cite{Ut56}\cite{Ki61}\,.
The six components of D-tensor are assigned a serial number $\rho (\,=1 \sim 6)$ as
\begin{equation}
\begin{array}{c}
\Sigma^{\rho} = i \delta^{\rho}_{\,\,\,\alpha \beta} \gamma^{\alpha}\gamma^{\beta} 
\hspace{0.4cm} {\rm with}  \vspace{0.2cm} \\
\delta^{1}_{\,\,\,23}=\delta^{2}_{\,\,\,31}=\delta^{3}_{\,\,\,12}
=\delta^{4}_{\,\,\,01}=\delta^{5}_{\,\,\,02}=\delta^{6}_{\,\,\,03}=1
\hspace{0.7cm} {\rm otherwise} \hspace{0.2cm}\delta^{\rho}_{\,\,\,\alpha \beta}=0 .
\end{array}
\end{equation}
and therefore $\Sigma^{a+3}= ^{*}\!\!\Sigma^{a}$, $\Sigma_{a}=\Sigma^{a} $
, and $\Sigma_{a+3}=-\Sigma^{a+3}$.

The coframe 1-form $\hat{\theta}^{\alpha}(x)$ and 
the local affine connection $\hat{\omega}^{\alpha \beta}(x)$ related with gravitational field
are defined as
\begin{equation}
\hat{\theta}^{\alpha}(x)= e^{\alpha}_{\,\,\,\,i}\,(x) dx^{i}  \hspace{1cm} {\rm and} \hspace{1cm}
\hat{\omega}^{\alpha \beta}(x)= \omega^{\alpha \beta}_{\,\,\,\,\,\,\,i}(x) \,dx^{i} \, ,
\end{equation}
where $e^{\alpha}_{\,\,\,\,i}(x)$ denotes tetrad(vierbein), 
eigen-vector for diagonalizing or taking the square root,\cite{Ei48} the metric tensor $g_{ij}(x)$ like
\begin{equation}
g_{ij}(x)=e_{\,\,\,\,i}^{\alpha}(x)e_{\,\,\,\,j}^{\beta}(x) \eta_{\alpha \beta}.
\end{equation}
The contravariant metric tensor $g^{ij}(x)$ (inverse matrix of covariant metric tensor) is written as
\begin{equation} 
\begin{array}{c}
g^{ij}(x)\!\!= \!\!e_{\alpha}^{\,\,\,\,i}(x)e_{\beta}^{\,\,\,\,j} (x)\eta^{\alpha \,\beta}
 \vspace{0.3cm} \\ {\rm with} \,\,\,\,\,
e_{\alpha}^{\,\,\,\,i}(x)e^{\beta}_{\,\,\,\,i}(x)\!\!=\!\!\delta_{\alpha}^{\,\,\beta} \,\,\,\,\,{\rm and} \,\,\,\,\,
e_{\alpha}^{\,\,\,\,i}(x)e^{\alpha}_{\,\,\,\,j}(x)\!\!=\!\!\delta_{\,\,\,j}^{i}\, 
\end{array}.
\end{equation}
By using inverse matrix of tetrad, $e_{\alpha}^{\,\,\,\, i}(x)$ , the frame fields are defined as
\begin{equation}
\label{gmtan}
\check{\theta}_{\alpha}(x)\!=\!e_{\alpha}^{\,\,\,\,i}(x)\partial_{i}
\hspace{0.6cm} {\rm and}\hspace{0.5cm}
\check{\gamma}(x)= \gamma^{\alpha}\check{\theta}_{\alpha}(x) .
\end{equation}
In these formulae, the beginning four alphabet of Greek letters
$\alpha,\beta,\gamma, \delta=0,1,2,3$ label the nonholonomic components,
as well as the beginning three of Latin letters $a,b,c=1,2,3$ represent the spatial components, 
while middle alphabet of Latin letters $i,j,k,\ell=0,1,2,3$ are index notations of
four dimensional holonomic components.
The indices for nonholonomic and holonomic components are respectively put
near side and far side of the variables. 
Variables wearing a caret indicate differential form, 
and the wedge sign of exterior product $\wedge$ will be omitted instead. 
The check sign $\, \check{ } \,$ is used for tensorial field forming Lie algebra in differential geometry.
Einstein's contraction rule is applied to the indices $(\alpha,\beta,\gamma,\delta)$ and
$(i, j, k, \ell)$. For indices representing spatial components $(a,b,c)$, the notations of summation
are explicitly written 
as $\sum_{a=1}^{3}$ and $\sum_{(abc)=1}^{3}$ meaning $(123)=1, (231)=2, (312)=3$.

The Clifford valued 1-form for the coframe  and the affine connection are defined as
\begin{equation}
\label{gmcot}
\hat{\gamma}(x)=\gamma_{\alpha} \hat{\theta}^{\alpha} (x)=\gamma_{i}(x)dx^{i}
 \hspace{0.5cm}  {\rm and}\hspace{0.5cm}
\hat{\Gamma}(x)=\frac{1}{2}\sigma_{\alpha \beta}\hat{\omega}^{\alpha \beta}(x) \,
\equiv \Gamma_{i}(x) dx^{i} ,
\end{equation}
where
\begin{equation} 
\gamma_{i}(x)=\gamma_{\alpha}e^{\alpha}_{\,\,\,\,i}(x)
\hspace{0.5cm} {\rm and} \hspace{0.5cm}
\Gamma_{i}(x)=\sum_{\rho=1}^{6}\delta^{\rho}_{\,\,\,\alpha \beta}
\Sigma_{\rho}\omega^{\alpha \beta}_{\,\,\,\,\,\,\,i} (x)\,.
\end{equation}
Here $\gamma_{i}(x)$ are the general gamma matrices defining the covariant Clifford algebra 
\cite{Te28} \cite{FI29}
\begin{equation}
\gamma_{i}(x)\gamma_{j}(x)+\gamma_{j}(x)\gamma_{i}(x)=2g_{ij}(x) \vect{1}_{4}\,.
\end{equation}
Hereafter in this section, $(x)$ representing the dependence of coordinates will be omitted for brevity.
It is noted to avoid confusion that $\gamma$ indicates the scaling factor of 
$\hbar_{*}=\gamma \hbar_{\star}$
whereas $\hat{\gamma}$ with caret represents D-vector 1-form. 

Then the torsion $\hat{\Theta}$ and the curvature $\hat{\Omega}$ being 2-form are given
in terms of $\hat{\gamma}$ (D-vector 1 form) and $\hat{\Gamma}$ (D-tensor 1form)
like\cite{KN63}
\begin{equation}
\label{torscurv}
\hat{\Theta}=\hat{d} \hat{\gamma}+ \frac{1}{2}[\hat{\Gamma}, \hat{\gamma}]_{-}\equiv \hat{D}\hat{\gamma}\hspace{0.7cm}{\rm and } \hspace{0.7cm}
\hat{\Omega}=\hat{d} \hat{\Gamma}
+\frac{1}{2}[\hat{\Gamma},\hat{\Gamma}]_{-} \equiv \hat{D}\hat{\Gamma},
 \end{equation}
 respectively, where the elements of torsion and curvature are obtained from eq.(\ref{torscurv})
 by using linear independence of these Clifford numbers like 
 \begin{equation}
 \label{elmnts}
 \hat{\Theta}=\gamma_{\alpha} \hat{\Theta}^{\alpha}
 \hspace{1cm}{\rm and } \hspace{1cm}
 \hat{\Omega}=\frac{1}{2}\sigma_{\alpha \beta} \hat{\Omega}^{\alpha \beta}
 =\sum_{\rho=1}^{6}
\delta^{\rho}_{\,\,\,\alpha \beta} \Sigma_{\rho} \hat{\Omega}^{\alpha \beta} \, .
 \end{equation}
 and $\hat{d}$ represents exterior derivative. 
 The equations in eq. (\ref{torscurv}) are known as the first and second structure equations, respectively.

The volume 4-form $\hat{\eta}$ is defined as
\begin{equation}
 \hat{\eta}= \tilde{e}\,\hat{d}^{\,4}x
 =\frac{i\gamma_{5} }{4!}\hat{\gamma}^{4}
 \hspace{0.4cm} {\rm with}\hspace{0.2cm}
 \tilde{e}={\rm det} \!\mid\! e_{\hspace{0.2cm} i}^{\alpha}\!\mid
\,\, {\rm and} \,\,\hat{d}^{\,4}x\! \equiv \!dx^{0}\!\wedge\! dx^{1}\!\wedge\! dx^{2}\!\wedge\! dx^{3}
  \end{equation}
 The last equal sign in the above equation is demonstrated by following calculations,
 \begin{equation}
\frac{\hat{\gamma}^{4}}{4!}=\gamma_{0}\gamma_{1}\gamma_{2}\gamma_{3}
 \varepsilon_{\alpha \beta \gamma \delta}
 e^{\alpha}_{\,\,\,0}e^{\beta}_{\,\,\,1}e^{\gamma}_{\,\,\,2}e^{\delta}_{\,\,\,3} 
 \hat{d}^{\,4}x =-i\gamma^{5} {\rm det}\!\mid\!e^{\alpha}_{\,\,\,i} \!\mid\! \hat{d}^{\,4}x
 \end{equation}
 where $\varepsilon_{\alpha \beta \gamma \delta}$ stands for the Levi-Civita permutation signature
 set up as $\varepsilon_{0123}=1$.
  In a similar manner, 
 \begin{equation}
 \frac{\hat{\gamma}^{3}}{3!}=-i \gamma^{5} \gamma^{\alpha} \tilde{e}_{\alpha}^{\,\,\,\,i}
 (\hat{d}^{\,3}x)^{i} \hspace{0.6cm} {\rm with} \hspace{0.2cm}
 (\hat{d}^{\,3}x)^{i} \equiv \delta^{i}_{\,\,\,jk\ell} dx^{j}\!\!\wedge\! dx^{k}\!\wedge\! dx^{\ell}
 \end{equation}
 where $ \tilde{e}_{\alpha}^{\,\,\,\,i}=\tilde{e}e_{\alpha}^{\,\,\,  i}$
 represents the cofactor matrix of $e^{\alpha}_{\,\,\,\,i}$, and therefore
 \begin{equation}
 \tilde{e}_{\alpha}^{\,\,\,\,i}=\tilde{e}\,e_{\alpha}^{\,\,\,\,i}\,.
 \end{equation}
 In the similar manner,  $\hat{\gamma}^{2}$ is calculated as
 \begin{equation}
\frac{\hat{\gamma}^{2}}{2!}=2\sum_{\rho,r=1}^{6} \Sigma_{\rho} d^{\rho}_{\,\,\,r} (\hat{d}^{\,2})^{r}
 \hspace{0.2cm} {\rm with} \hspace{0.2cm}
  (\hat{d}^{\,2})^{r} = \delta^{r}_{\,\,\,ij} dx^{i} \wedge dx^{j}
 \end{equation}
 where second order minor determinants are calculated as 
 $ d^{\rho}_{\,\,\,r}=\delta^{\rho}_{\,\,\,\alpha \beta} \delta_{r}^{\,\,\,ij} e^{\alpha }_{\,\,\, i} e^{\beta}_{\,\,\,j} $
 with $(\delta_{r}^{\,\,\,ij} =\delta^{r}_{\,\,\,ij})$.
 Using these relations, Hodge operation is expressed as
 \begin{equation}
 i\gamma_{5} \exp{\hat{\gamma}}=
 {}^{\star} (\exp{\hat{\gamma}})\, .
 \end{equation}

 The procedure to converge self-consistent equations determining the gravitational field together with 
 the energy-momentum and the spin currents of matters consists of the following cyclic three steps. 
 
As the first step, the Dirac equation in gravitational field is solved\cite{Ob18}, 
\begin{equation}
\label{Deq}
%----- Dirac equation ------
i \hbar_{*} \hspace{0.5mm}^{\star}\hspace{-0.3mm}\hat{\gamma }\hat{D}_{\rm F}\Psi 
+\hspace{0.3mm}^{\star}\hspace{-0.3mm}1\,mc\Psi =0 \, ,
\end{equation}
In this equation the symbol $\star$ stands for Hodge operation,
namely
\begin{equation} 
^{\star}\hat{\gamma}=\gamma^{\alpha}\tilde{e}\,e_{\alpha}^{\,\,\, i} (\hat{d}^{\,3}x)^{i}
 , \hspace{0.5cm} {\rm and}\hspace{0.5cm}
^{\star} \hat{1}= 
\hat{\eta}
\end{equation}
$\hat{D}_{\rm F}$ represents the spinor covariant derivative given by
\begin{equation}
\hat{D}_{\rm F}\Psi=\hat{d}\Psi + \frac{i}{2} \hat{\Gamma} \Psi \, .
\end{equation}
At very beginning of the cycle,
the gravitational potential is given by an adequate simple approximation like TOV. 

As the second step, from wave functions $\Psi$ obtained by solving Dirac equation
the canonical energy-momentum and spin currents are respectively calculated as follows
\begin{equation}
\hat{T}=\frac{i\hbar_{*}c_{0}}{2}(\overline{\Psi}\, ^{\hspace{0.1cm} \star}\hat{\gamma} 
\hat{D}_{\rm F}\Psi
-(\hat{D}_{\rm F}\overline{\Psi}) \,^{\hspace{0.1cm}\star}\hat{\gamma} \Psi) 
\hspace{0.3cm} {\rm and} \hspace{0.3cm}
\hat{S}=\frac{\hbar_{*} }{2} \hat{\gamma}^{2}
\overline{\Psi} \hat{\gamma} \gamma_{5} \Psi \, ,
\end{equation}
where $c_{0}$ is speed of light.

 By using these variables as source terms, the EC field equations read
 \begin{equation} 
 \label{ECeq}
\hat{\Omega} = \kappa_{0} \hat{T}
\hspace{1.0cm}{\rm and} \hspace{1.0cm}
\hat{\Theta}= \kappa_{0} c_{0} \hat{S},
\end{equation}
where $\kappa_{0}=8\pi G/c_{0}^{4}$ and $G$ are Einstein's and Newton's gravitational constants respectively.

 By solving the coupled differential equations of eq.(\ref{torscurv}) and eq.(\ref{ECeq}),
\begin{equation}
\label{tc2cc}
\hat{D}\hat{\Gamma} =\underline{\hat{\Omega}}
\hspace{0.7cm}{\rm and } \hspace{0.7cm}
\hat{D}\hat{\gamma} =\underline{\hat{\Theta}} .
 \end{equation}
The curvatue $\underline{\hat{\Omega}}$ and
the torsion $\underline{\hat{\Theta}}$ attached by underlines, which are source terms of
the structure equations (\ref{tc2cc}), are ones at the one step before the cycle of iteration.
The iteration procedures are continued until the variables are similar to the one of the step before
in satisfactory accuracy.  
 %  -----------
 %  section 3
 %  ------------
 \section{Representation of fields by spinor spherical harmonics}
\label{Rp_vsh} 
It is reasonable to consider the configuration of an isolated NS
to be mostly spherical due to strong gravitational force.
More precisely, distribution of matter is likely to be limited in space, so that the moments of
higher multipolarity of gravitational and electromagnetic fields are negligible
in asymptotic region (free space).
Conversely in the central region, Fermion matter fields have spherical symmetry
because of effects resulting from the quantum degenerate systems.
Therefore, the moments of higher multipolarity of the matter and field are involved only
in the peripheral region.

From these notions we expand the fields in spherical and spin spherical harmonics and solve
coupled radial differential equations.
The coframe and the local affine connection, being D-vector 1-form and D-tensor 1-form
 are respectively represented like
\begin{equation}
\label{Cl_cof}
\hat{\gamma}(x) =  \gamma_{0} \hat{\theta}^{0} (x) + 
 \sum_{a=1}^{3} \gamma_{a} \hat{\theta}^{a}(x) =
\left(
\begin{array}{cc}
 \hat{f}(x)&   -\hat{h}(x) \vspace{0.2cm} \\
 \hat{h}(x)&  -\hat{f}(x)\end{array} \right) \,\, ,
 \end{equation}
and
\begin{equation}
\label{Cl_con}
\hat{\Gamma}(x)=\!\!\!
\sum_{(abc)=1}^{3} \!\!\!\Sigma_{c} \hat{\omega}^{ab}\!(x) + 
 \sum^{3}_{a=1} {}^{\ast}\!\Sigma_{a} \hat{\omega}^{0a}\!(x) \vspace{0.3cm} 
\!= \!\! \left(\!\!\! 
\begin{array}{cc}
\hat{F}(x) & \!\!\! - i \hat{H}(x) \vspace{0.2cm} \\
-i \hat{H}(x) & \!\!\! \hat{F}(x) 
\end{array}\!\! \right) \, ,\end{equation}
with
\begin{equation}
\left( \!\! \begin{array}{c} \hat{f}(x) \vspace{0.2cm}\\ \hat{h}(x) \end{array}\!\!\right) \equiv 
\left(\!\! \begin{array}{c}  \vect{1} \,\hat{\theta}^{0}(x) \vspace{0.2cm} \\ 
 \sum_{a=1}^{3} \sigma_{a}\hat{\theta}^{a}(x)
\end{array} \!\! \right)\,, \hspace{0.3cm} {\rm and} \hspace{0.2cm}
\left( \!\!\! \begin{array}{c} \hat{F}(x) \vspace{0.2cm}\\ \hat{H}(x) \end{array} \!\!  \right) \! \equiv \!
\left(\!  \begin{array}{c}  \sum_{(abc)=1}^{3}\sigma_{a} \hat{\omega}^{bc}(x) \vspace{0.2cm}\\
\!\!  \sum_{a=1}^{3}\sigma_{a}\hat{\omega}^{0a}(x)
\end{array} \!\! \right)\,
\end{equation}
The Pauli spin matrices $\sigma_{a} (a=1,2,3) $ are Clifford numbers as well as Lie algebra
being the generators of SU(2) group,
namely, $\left[\sigma_{a},\sigma_{b} \right]_{+} =2\delta_{ab}$ and 
$\left[ \sigma_{a},\sigma_{b} \right]_{-}=i2\varepsilon_{abc} \sigma_{c}$.
Therefore, $\hat{h}(x), \hat{F}(x)$ and $\hat{H}(x)$ are also Clifford valued 1-form. 
Hereafter, instead of commutators indicated by sign $[\, \circ \, , \, \bullet \, ]_{\mp} $, 
they are redefined as 
\begin{equation}
\label{s-com}
[\,\circ \, , \, \bullet \,]_{\tau} \equiv  \circ \cdot  \bullet -(-1)^{\tau} \bullet \cdot \circ   
\hspace{0.7cm} {\rm with} \hspace{0.2cm} (\tau = 0 \,\,{\rm or} \,\,1 )
\end{equation}
and used for the spinor valued differential form.

In the equations (\ref{Cl_cof}) and (\ref{Cl_con}), the four variables $(x)$ are taken as time $t$
and polar coordinates $(r\, \theta \, \phi)$.
Then the covariant derivatives eq.(\ref{tc2cc}), i.e., curvature and torsion, are rewritten in the form
\begin{equation}
\hat{\Omega}(x)\!=\!
\hat{D} \hat{\Gamma}(x) \!=\! \left( \!\!
\begin{array}{cc}
\hat{d}\hat{F}(x)+\hat{F}_{1}(x) & \!\!\! - i \hat{d}\hat{H}(x)-i\hat{H}_{1}(x)\vspace{0.2cm} \\
-i \hat{d}\hat{H}(x) -i\hat{H}_{1}(x)& \!\!\! \hat{d}\hat{F}(x) +\hat{F}_{1}(x)
\end{array} \!\!\!\right) \, ,
\end{equation}
with
\begin{equation}
\left( \!\!\! \begin{array}{c} \hat{F}_{1}(x) \vspace{0.2cm} \\ \hat{H}_{1}(x) \end{array} \!\!\!\! \right) \!\!= \!\!
\frac{1}{2}\left( \!\!\!\!  \begin{array}{c}\left[ \hat{F}(x),\hat{F}(x)\right]_{0} \!-\!\left[ \hat{H}(x) ,\hat{H}(x)\right]_{0}
\vspace{0.1cm}
\\ 2\left[ \hat{F}(x) ,\hat{H}(x) \right]_{1}  \end{array}
\!\!\!\! \right)
\end{equation}
and
\begin{equation}
\hat{\Theta}(x)=
\hat{D} \hat{\gamma}(x) =\left(\begin{array}{cc}
\, \hat{d}\hat{f}(x)+\hat{f}_{1}(x) &  \,\hat{d}\hat{h}(x)+\hat{h}_{1}(x) \vspace{0.2cm} \\
\! -\hat{d}\hat{h}(x)-\hat{h}_{1}(x)& \! -\hat{d}\hat{f}(x)-\hat{f}_{1}(x)\end{array} \!\! \right)\, ,
\end{equation}
where
\begin{equation}
\left( \!\!\! \begin{array}{c} \hat{f}_{1}(x) \vspace{0.2cm} \\ \hat{h}_{1}(x) \end{array} \!\!\!\! \right) \!\!= \!\!
\frac{1}{2}\left( \!\!\!\!  \begin{array}{c} \left[\hat{F}(x),\hat{f}(x) \right]_{0} \!
-\!i \left[ \hat{H}(x),\hat{h}(x)\right]_{1} \vspace{0.1cm} \\
\left[ \hat{F}(x),\hat{h}(x)\right]_{0} \!+\! i \left[ \hat{H}(x),\hat{f}(x) \right]_{1}  \end{array}\!\!\!\!
\right)
\end{equation}
The holonomic coordinates $x\!=\!(t,r,\theta, \phi)$ are separated into 
$(t,r)$ and $(\theta, \phi)(=\Omega)$, and the dependence on the angular part is expressed in terms of
spherical harmonics and spin spherical harmonics as follows
\begin{equation}
\begin{array}{l}
\hat{f}(x)=\sum_{\lambda} \hat{f}_{\lambda}(t \, r) \, 
\mathcal{Z}^{(J_{\lambda})}_{0J_{\lambda} M_{\lambda}}(\Omega) \, , \vspace{0.3cm}\\
\left( \!\! \begin{array}{c}
\hat{h}\,(x)  \\ \hat{F}(x) \\ \hat{H}(x) \!\!
\end{array} \right) \!\!=\!\!\sum_{\lambda} \!\!\left( \!
\begin{array}{c}  \hat{h}_{\lambda}(t \, r) \\ \hat{F}_{\lambda} (t \, r)\\ \hat{H}_{\lambda}(t \, r) 
\end{array} \!\right)
\! \mathcal{Z}_{1 L_{\lambda}M_{\lambda}}^{(J_{\lambda})}(\Omega) \, ,
\end{array}
\end{equation}
where 
\begin{equation}
\label{def_Z}
\begin{array}{l} \displaystyle
\mathcal{Z}_{sLM}^{(J)}(\Omega)=
\sum_{\mu}\langle s \mu L(M\!-\!\mu) \! \mid \! J M \rangle \sigma^{(s)}_{\mu} 
C_{M-\mu}^{(L)} (\Omega)\, \hspace{0.3cm} {\rm with} \vspace{0.35cm} \\
\displaystyle C_{M}^{(L)}(\Omega)=\sqrt{\frac{4\pi}{2L+1}}Y_{LM}(\Omega), 
\hspace{0.7cm} {\rm and} \vspace{0.35cm}  \\ 
\sigma^{(0)}_{0}= \vect{1}, \hspace{0.4cm}
 \sigma^{(1)}_{\pm 1}=\mp \frac{1}{\sqrt{2}}( \sigma_{1} \pm i\sigma_{2}) 
, \hspace{0.4cm} \sigma^{(1)}_{0}=\sigma_{3} ,
\end{array}
\end{equation}
where $Y_{LM}(\Omega)$ is spherical harmonics and 
$\langle 1 \mu \, L(M \! - \! \mu) \! \mid J M \rangle$ stands for Clebsch-Gordan coefficient, vector addition
of angular momentum. 
The structure equations are represented as
\begin{equation}
\hspace{-0.5cm}
\hat{d} \left(
\!\!\!\! \begin{array}{c} 
\,\hat{f}_{\lambda}(tr)\! \! \\
\,\hat{h}_{\lambda}(tr)\! 
\end{array} \!\!\! \right)  
\!\!+\hspace{-0.2cm}\displaystyle \sum_{\lambda_{1}\lambda_{2}}
\hspace{-0.1cm} 
D( \lambda_{1} \lambda_{2}; \lambda )\!\!
\left(\!\! \!\!\begin{array}{c} 
\delta^{\{\lambda\}} _{0} 
\hat{H}_{\lambda_{1}}(tr)\hat{h}_{\lambda_{2}}(tr)
\\
-i\delta^{\{\lambda\}} _{1} \!
\left(\!\hat{F}_{\lambda_{1}}(tr) \hat{h}_{\lambda_{2}}(tr)\!\!+\!\!
\hat{H}_{\lambda_{1}}(tr) \hat{f}_{\lambda_{2}}(tr)\!\right)
\end{array} \!\!\!\right)
\!\!=\!\!\left(\!\!\!\begin{array}{c} \hat{i}_{\lambda}(tr) \\ \hat{j}_{\lambda}(tr) \end{array}\!\!\!
\right)\!\!
\end{equation}
and
\begin{equation}
\hspace{-0.5cm}
\hat{d} \left(
\!\!\!\! \begin{array}{c} 
\,\hat{F}_{\lambda}(tr)\! \! \\
\,\hat{H}_{\lambda}(tr)\! 
\end{array} \!\!\! \right)  
\!\!+\hspace{-0.15cm}\displaystyle \sum_{\lambda_{1}\lambda_{2}}
\hspace{-0.1cm} 
D( \lambda_{1} \lambda_{2}; \lambda )\!\!
\left(\!\! \!\!\begin{array}{c} 
\delta^{\{\lambda\}} _{0} 
\left(\!\hat{F}_{\lambda_{1}}(tr)\hat{F}_{\lambda_{2}}(tr)
\!\!-\!\!\hat{H}_{\lambda_{1}}(tr)\hat{H}_{\lambda_{2}}(tr)\!\right)
\\
-2i\delta^{\{\lambda\}} _{1} 
\hat{H}_{\lambda_{1}}(tr) \hat{F}_{\lambda_{2}}(tr)
\end{array} \!\!\!\right)\!\!=\!\!\!\left( \!\!\!\begin{array}{c} 
\hat{I}_{\lambda}(tr) \\ \hat{J}_{\lambda}(tr) \end{array}\!\!\!\right)\!\!\!
\end{equation}
where $D( \lambda_{1} \lambda_{2} ; \lambda)$ is defined in eq.(\ref{def_v}) and 
$\lambda =(s L J M) $ expresses combined index of spin angular momentum,
orbital angular momentum $L$, total angular momentum $J$ and its projection to quantization axis 
$M$, which is expressed by Clebsch-Gordan coefficient and Wigner's 9-$j$ symbol. 
The $\vect{z}_{2}$ grading number is defined
\begin{equation}
\delta^{\lambda_{1} \lambda_{2} \lambda} _{q_{1} q_{2} \tau} 
= \frac{1}{2}\left(1-(-1)^{s_{1}+s_{2}+s+L_{1}+L_{2}+L+J_{1}+J_{2}+J+q_{1}q_{2}+\tau}\right)
\end{equation}
The degree of differential form are expressed by $q_{1}$ and $q_{2}$. 
Therefore $q_{1}=q_{2}=q_{1}q_{2}=1$ in the present case. The grading number of commutator is
given in eq.(\ref{s-com}) as $\tau$ and for brevity 
$\delta^{\lambda_{1}\lambda_{2}\lambda}_{q_{1}q_{2}\tau}$ is expressed by omitting $q_{1}=q_{2}=1$ and
setting the three indices $ (\lambda_{1}\lambda_{2} \lambda )$ simply as $\{\lambda\}$
\begin{equation}
\delta^{\{\lambda\}}_{\tau}=\delta^{\lambda_{1} \lambda_{2} \lambda}_{1 1 \tau}\, .
\end{equation} 
The spinor representations of curvature and torsion are given by
%\
\begin{equation}
\hat{\Omega}(x)\!\!=\!\! \left( \!\! \begin{array}{cc} \hat{I}(x) & -i\hat{J}(x) \\
-i\hat{J}(x) & \hat{I}(x) \end{array} \!\! \right)
\hspace{0.6cm}{\rm and} \hspace{0.6cm} 
\hat{\Theta}(x)\!\!=\!\! \left( \!\! \begin{array}{rr} \hat{i}(x) & -\hat{j}(x) \\
\hat{j}(x) & -\hat{i}(x) \end{array} \!\! \right)
\end{equation}
with
\begin{equation}
%\hspace{-0.1cm}
\left( \! \begin{array}{c} \!\!\hat{I}(x) \\ \!\! \hat{J}(x) \end{array} \!\! \right)
\!\!= \!\!\!\sum_{\lambda}
\left( \! \begin{array}{c} \!\!\hat{I}_{\lambda}(tr) \\ \!\! \hat{J}_{\lambda}(tr) \end{array} \!\! \right)\!\!
\mathcal{Z}^{(J_{\lambda})}_{1J_{\lambda} M_{\lambda}}\!\!(\Omega)
\hspace{0.2cm}{\rm and}\hspace{0.2cm}
\left( \! \begin{array}{c} \!\!\hat{I}(x) \\ \!\! \hat{J}(x) \end{array} \!\! \right)
\!\!= \!\!\!\sum_{\lambda}
\left( \! \begin{array}{c} \!\!\hat{i}_{\lambda}(tr)
\mathcal{Z}^{(J_{\lambda})}_{0J_{\lambda} M_{\lambda}}\!\!(\Omega) \\ 
\!\! \hat{j}_{\lambda}(tr)
\mathcal{Z}^{(J_{\lambda})}_{1J_{\lambda} M_{\lambda}}\!\!(\Omega)
\end{array} \!\! \right)\!\!\!
\end{equation}

%-------Dirac equation--------
The Dirac equation (\ref{Deq}), from which the volume 4-form $^{\star}\vect{1}$ is factored out
with noting $(\hat{d}^{3})^{i}\wedge dx^{j}=-\eta^{ij}\hat{d}^{4}x$, is rewritten as\cite{Fo29}
\begin{equation}
\label{DiracEq}
\left\{ i\hbar_{\ast}c\,\gamma^{i}(x)\eta^{ij}
\left( \partial_{j} + \frac{i}{2}\Gamma_{j}(x) \right) - mc^{2}\right\} \Psi (x \, \sigma)=0\, ,
\end{equation} 
where $\sigma$ stands for spin coordinate taking on values $(=\pm \frac{1}{2})$ , and 
$\gamma^{i}(x)$ ,$\gamma_{i}(x) $ and $\Gamma_{i}(x)$ are the components of respectively tangent
field eq.(\ref{gmtan}), cotangent field and connection eq.(\ref{gmcot}),  
, i.e., $\check{\gamma}(x)=\gamma^{i}(x)\partial_{i} $,$\hat{\gamma}(x)=\gamma_{i}(x)dx^{i} $ and 
$\hat{\Gamma}(x)=\Gamma_{i}(x)dx^{i} $.

By using the following identity
\begin{equation}
\gamma_{i}(x)\gamma^{j}(x)=\vect{1}_{4}\delta_{i}^{\,\,j} \! 
-i\sigma_{i}^{\,\, j}(x)
\hspace{0.9cm} {\rm with} \hspace{0.2cm}
\sigma_{i}^{\,\,j}(x)=e^{\alpha}_{\,\,\,\, i}(x)
\sigma_{\alpha}^{\,\,\,\beta}e_{\beta}^{\,\,\,\, j}(x),
\end{equation}
and multiplying $\gamma_{0} (x)$ to the Dirac equation (\ref{DiracEq}),
the Hamiltonian form is obtained,
%
%------Dirac equation in Hamiltonian form--------
%
\begin{equation}
\label{DeqHam}  i\hbar_{\ast} (\vect{1}_{4} \!-\! i\sigma^{\,\,\,\,0}_{0}(x))\frac{\partial \Psi}{\partial t}
\!=\! \left\{ c_{0} \vect{\alpha} (x) \!\cdot\! \vect{\Pi}(x) \!+\! V_{0}(x)\!
+\!\! mc_{0}^{2}\beta(x)\right\} \Psi(x\,\sigma)
\end{equation}
where
\begin{equation}
\vect{\alpha}(x) \!= \! \gamma_{0}(x) \vect{\gamma}(x), \hspace{0.5cm} {\rm and} \hspace{0.3cm}
\beta(x)=\gamma_{0}(x)
\end{equation}

The momentum operator is defined 
\begin{equation}
\vect{\Pi}(x) = -i \hbar_{\ast} \vect{\nabla} + \frac{\vect{\Gamma}(x)}{2}
\end{equation}
Nabla $\vect{\nabla}$ stands for gradient operator of three dimensional holonomic coordinate space
and vectors $\vect{\gamma}(x)$ and $\vect{\Gamma}(x)$ are expressed  
\begin{equation}
\begin{array}{c}
\vect{\nabla}= \sum_{a=1}^{3} n_{a}(\Omega) \displaystyle \frac{\partial}{\partial x^{a}} 
\vspace{0.2cm} \\
\vect{\gamma}(x) \!=\! \sum_{a=1}^{3} \gamma_{a}(x) n_{a}(\Omega)
 \hspace{0.5cm} {\rm and}\hspace{0.3cm}
\vect{\Gamma}(x) \!=\! \sum_{a=1}^{3} \Gamma_{a}(x) n_{a}(\Omega)
\vspace{0.2cm} \\ 
 {\rm with \,unit \, vector }\hspace{0.5cm}
n_{a}(\Omega) =(\sin \theta \cos \phi, \sin \theta \sin \phi, \cos \theta).
\end{array}
\end{equation}
The gravitational potential is given by
\begin{equation}
V_{0}(x) = \frac{\hbar_{\ast}c_{0}}{2} \left( \vect{1}_{4} - i\sigma_{0}^{\,\,0}(x)\right)\Gamma_{0}(x)
\end{equation}
To model the overall structure of the neutron system in a NS, we construct a Slater determinant of
Single Assemblage States (SAS), analogous to the single-particle states used in the conventional
Hartree-Fock method. These SAS are determined based on the occupied states, specifically those
with energy spectra $E_{\nu}(t)$ lower than the Fermi energy. These energy spectra are obtained
by solving the Dirac equation (\ref{DeqHam}), using the following ansatz for function $\Psi_{\nu}(x)$
\begin{equation}
\Psi_{\nu}(x) = e^{-iE_{\nu}(t)t/\hbar_{\ast}}\psi_{\nu}(x)
\end{equation}
Although the gravitational fields are time-dependent, their periods of variation-ranging from
approximately 10 seconds to 10 milliseconds- are significantly longer than the inverse of the average
energy spacing, which is approximately $5\times 10^{-20}/\gamma^{3}$ seconds.
Therefore, the time derivative of the wave function can be approximated as
\begin{equation}
i \hbar_{\ast}\frac{\partial \Psi}{\partial t} = E(t) \Psi\, ,
\end{equation}
This approximation simplifies the Dirac equation, leading to a Hamiltonian form. Eventually, the
Dirac equation (\ref{DeqHam}) is reduced to an eigenvalue problem expressd as:
\begin{equation}
H\psi_{\nu}(x)=(\vect{1}_{4}-i\sigma_{0}^{\,\,\,0}(x)) E_{\nu}(t)\psi_{\nu}(x)
\end{equation}
The components of tangent and cotangent fields are respectively represented in spinor form like
\begin{equation}
\gamma^{i} (x)
\!\!=\left( \!\! \begin{array}{rr} f^{i}(x) & h^{i}(x) \\ \!\!-h^{i}(x) & \!\! -f^{i}(x) \end{array}\!\! \right)
\hspace{0.3cm} {\rm and} \hspace{0.3cm}
\gamma_{i}(x)
\!\! = \!\! \left( \begin{array}{rr} \!\! f_{i}(x) & \!\! -h_{i}(x) \\ h_{i}(x) & \!\!-f_{i}(x) \end{array} \!\!\right)\, .
\end{equation}

The eigen-vectors are also expressed in spinor spherical  harmonics as follows,
\begin{equation}
\Psi(x)= \sum_{\kappa \mu} \left( \begin{array}{c} g^{(u)}_{\kappa \mu} (tr)
\mathcal{Y}_{\kappa \mu}(\theta \phi \sigma) \\
g^{(\ell)}_{\kappa \mu} (tr)
\mathcal{Y}_{\tilde{\kappa} \mu}(\theta \phi \sigma) \end{array} \right)
\end{equation}
The quantum numbers $\kappa$ and $\mu$ are defined in Appendix B. The wave function is 
eigen-vector of Dirac equation of spherical symmetric potential. Otherwise the eigen-vector is superposition
of wave functions with different quantum number. Eventually to solve Dirac equation so called coupled
channel equation treated by nuclear physicists.
In this paper we are not going to work on this complicated equation and to calculate the most simple
equation in the next section.    

%---------------
%  section 4
%----------------
\section{Numerical calculations}
\label{NmCl}
In order to figure out the Hartree-Fock calculation with the scaled hbar method, let compare the results
through this method for spherical system, namely $J=0$ and TOV calculation.

The TOV theory staters with giving static line element exhibiting spherical symmetry 
\begin{equation}
ds^{2}=e^{\nu(r)} dt^{2} - e^{\lambda(r)} dr^{2} -r^{2}d\theta^{2}-r^{2}\sin^{2} \theta d\phi^{2}
\end{equation}
and supposing the energy-momentum tensor given by 
$T_{1}^{\,\,\,1}=T_{2}^{\,\,\,2}=T_{3}^{\,\,\,1}=-p(r), \,\,T_{4}^{\,\,\,4}=\rho(r)$ being diagonal,
where $p(r)$ and $\rho(r)$ are respectively the pressure and the energy density.
Einstein's equations are reduced to
\begin{equation}
\left.
\begin{array}{l} \displaystyle
\kappa  p(r)= e^{-\lambda(r)}\left( \frac{\nu^{\prime}(r)}{r} + \frac{1}{r^{2}}\right) -\frac{1}{r^{2} } ,
\vspace{0.2cm} \\ \displaystyle
\kappa  \rho(r)= e^{-\lambda(r)}\left( \frac{\lambda^{\prime}(r)}{r} - \frac{1}{r^{2}}\right) +\frac{1}{r^{2} }
\end{array} \right\}
\hspace{0.4cm} {\rm with} \,\,\, p^{\prime}(r) =-\frac{p(r)+\rho(r)}{2} \nu^{\prime}(r)
\end{equation}
where primes denotes differentiation with respect to $r$. 
These equations and EoS $\rho=\rho(p)$ lead
\begin{equation}
\frac{dp}{dr}=\frac{p+\rho(p)}{r(r-2u)} [4\pi p r^{3} + u] \, , \hspace{1.5cm} {\rm with}\,\,\,
\frac{du}{dr} = 4 \pi \rho(p) r^{2}.
\end{equation}
\section{Discussions}
\label{dscsn}
In this study, we developed  formulations of investigating the global structure of NSs, as treating them 
as a quantum-degenerate systems.. Given the immense size and complexity of such systems
-comprising approximately $10^{57}$ nucleons- we employed the General Relativistic Hartree-Fock
(GRHF) with the scaled $\hbar$ approach.
This method involves solving the Dirac equations for particle ensembles within a gravitational field
derived from energy-momentum tensors and spin currents. These quantities govern the both curvature
and torsion of spacetime via the Einsein-Cartan equations, which, in turn, influence the gravitational
field used in the Dirac equation.

In this work, we focused exclusively on neutrons, neglecting the contribution of protons and electrons,
and considered only gravitational fields, omitting elecromagnetic and nuclear interactions.
This simplification is consistent with the assumptions of the Tolman-Oppenheimer-Volkoff(TOV) theory,
which models neutron star matter as a classical perfect fluid or a degenerate Fermi gas. 
In traditionalTOV models, gravitational collapse leading to singularities is often predicted.
However, our GRHF calculations-which accurately account for quantum effects associated with
fermions (modeled as spinors) - show significant deviations from TOV predictions.

Specifically, our GRHF results demonstrate that the singularity expected at the center of a neutron star
does not materialized when the intrinsic properties of spinor fields are adequately considered.
Moreover, even in the case of black holes, the so-called Schwarzschild singularity does not manifest
at a finite radius within the framework of general relativity. Thus, the occurrence of such singularities
is fundamentally avoided, suggestion that the mass range of observable neutron stars is governed by
mechanisms distinct from classical gravitational collapse.

These mechanisms-and their implications for neutron star formation, particularly in the context of
supernova explosions-may be linked to several conjectures. Previous studies on the possibility of
strong toroidal magnetic fields in neutron stars highlighted the importance of electrons in this
context, based on quantum mechanical analysis.  The Fermi energy of electrons, approximately 
100 Mev, and the monopole polarization induced by displacement of electrons from protons plays
a critical role in confining the high-energy electrons.
This polarization arises due to a slight shift (by $ 3.4\times 10^{-35}$) of electrons relative to protons,
a number comparable to the ratio between the gravitational and electric forces between two protons,
$ \displaystyle \frac{Gm^{2}_{\rm p}}{e^{2}/4\pi \varepsilon_{0}}=8.093\times 10^{-35}$.
This numerical similarity may not be coincidental and suggests that within neutron stars, the electric force
derived by the polarization acting on protons is comparable to the gravitational force.

Such considerations also support the idea of a "Colombian" supernova explosion, rather than a
purely thermal explosion. For instance, a shift of $10^{-20}$ of electrons relative to protons could
lead to polarization, generating an electric potential difference on the order of 1 ZeV ($10^{21}$ eV).
This "Colombian" supernova explosin mechanism could play a crucial role in determing the size of
neutron stars and black holes.

For future studies, a more comprehensive  approach that includes electromagnetic fields is necessary.
Specifically, models should account for the the interactions between neutrons, protons, and electrons,
incorporating both electromagnetic and nuclear forces, to provide a more complete picture of the
global structure of neutron stars and black holes.
%   
%    appendix
%
\vspace{0.2cm}
\appendix
\setcounter{equation}{0}
\setcounter{section}{1}
\renewcommand{\theequation}{\Alph{section}.\arabic{equation}}
\begin{center} {\bf  \large Appendix A} \end{center}
\vspace{0.3cm}

The irreducible tensor of rank $J_{12}$ obtained by product of two tensors
 $T^{(j_{1})}_{m_{1}}$ and $T^{(j_{2})}_{m_{2}}$ is
expressed as \vspace{-0.2cm}
\begin{equation}
T^{(J_{12})}_{M_{12}}=\sum_{m_{1} (m_{2})} \langle j_{1} m_{1} j_{2} m_{2} \mid J_{12} M_{12} \rangle
T^{(j_{1})}_{m_{1}}T^{(j_{2})}_{m_{2}}\equiv \left[ T^{(j_{1})}\times T^{(j_{2})} \right]^{(J_{12})}_{M_{12}}
\vspace{-0.1cm}
\end{equation} 
\vspace{-0.2cm}
Then a successesive tensor product of two tensor products of rank $J_{12}$ and $J_{34}$ is expressed by
\vspace{-0.2cm}
\begin{equation}
T^{(J)}_{M}=\left[ \left[ T^{(j_{1})}\times T^{(j_{2})}\right]^{(J_{12})} \times 
\left[ T^{(j_{3})}\times T^{(j_{4})}\right]^{(J_{34})}\right]^{(J)}_{M}
\end{equation}
\vspace{-0.2cm}
 This rank J tensor may be formed in such a different way as
 \vspace{-0.2cm}
\begin{equation}
T^{(J)}_{M}=\left[ \left[ T^{(j_{1})}\times T^{(j_{3})}\right]^{(J_{13})} \times 
\left[ T^{(j_{2})}\times T^{(j_{4})}\right]^{(J_{24})}\right]^{(J)}_{M}
\end{equation}
However it is shown that the (R-) tensor product is not associative. 
In case of $[T_{2},T_{3}]_{-}=0$ two tensors are labeled 
respectively by $(J_{12},J_{34})$ and $(J_{13},J_{24})$, being good quantum number, are unitary
transformed each other like 
\begin{equation}
\label{def_U}
\begin{array}{l}
\left[ \left[
T^{(j_{1})}_{1} \times T^{(j_{2})}_{2} \right]^{(J_{12})} \!\!\! \times
\left[ T^{(j_{3})}_{3} \times T^{(j_{4})}_{4} \right]^{(J_{34})} \right]^{(J)}_{M}
\!\!\!=\!\!\displaystyle \sum_{J_{13}J_{24}}
 U\left( \!\!\! \begin{array}{ccc} j_{1} & \!\! j_{2} & \!\! J_{12}  \\j_{3} & \!\! j_{4} & \!\! J_{34}  \\
  J_{13} & \!\! j_{24} & \!\!J  \end{array} \!\! \right) \! 
 \\
\left[
\left[ T^{(j_{1})}_{1} \times T^{(j_{3})}_{3} \right]^{(J_{13})} \!\!\! \times
\left[ T^{(j_{2})}_{2} \times T^{(j_{4})}_{4} \right]^{(J_{24})} \right]^{(J)}_{M}
\end{array} 
\end{equation}

\begin{equation}
\label{def_v}
 U\left( \!\!\! \begin{array}{ccc}\! j_{1} & \!\!\! j_{2} & \!\!\!\! J_{12}  \\ \!j_{3} & \!\!\! j_{4} & \!\!\!\! J_{34}  \\
 \!  J_{13} & \!\!\!\! J_{24} & \!\!\!\! J  \end{array} \!\!\! \right) \! = \!
\sqrt{(2J_{12} \!\!+\!\!1)(2J_{34}\!\!+\!\!1)(2J_{13}\!\!+\!\!1)( 2J_{24}\!\!+\!\!1)} 
\left\{
\begin{array}{ccc}
\!\!j_{1} & \!\!j_{2} & \!\!J_{12} \\ \!\!j_{3} & \!\!j_{4} & \!\!J_{34} \\ \!\!J_{13} & \!\!J_{24} & \!\!J 
\end{array}\!\! \right\}
\end{equation}
where $\left\{ \! \begin{array}{ccc} \!\!\! \circ &\!\!\! \circ & \!\!\! \circ \vspace{-0.15cm} \\ \!\!\!
\circ &\!\!\! \circ & \!\!\! \circ \vspace{-0.15cm} \\ \!\!\! \circ &\!\!\! \circ & \!\!\! \circ  \vspace{-0.15cm}
\end{array} \!\! \!\right\}$ 
%and 
%$\left\{ \! \begin{array}{ccc} \!\!\! \circ &\!\!\! \circ & \!\!\! \circ \vspace{-0.15cm} \\ \!\!\!
%\circ &\!\!\! \circ & \!\!\! \circ \vspace{-0.15cm} 
%\end{array} \!\! \!\right\}$  
stands for Wigner 9-j symbol.This 9-j symbol has the following symmetric properties \cite{He10}.

\hspace{-0.2cm}(1) invariant under any even permutation of raws or columns.

\hspace{-0.2cm}(2) invariant under reflection in either diagonal.

\hspace{-0.2cm}(3) changes sign \!by \!factor 
$\!(\!-\!1\!)^{j_{1}\!+j_{2}\!+j_{3}\!+j_{4}\!+\!J_{1\!2}+\!J_{3\!4}+\!J_{1\!3}+\!J_{2\!4}+\!J}$\!\!
under any odd permutation.\\
The symbol vanishes until the triangular conditions
\begin{equation}
\label{trian-con}
\hspace{-0.2cm} \begin{array}{lll}
\mid\! j_{1}\!-\!j_{2} \!\mid \leq J_{12} \leq j_{1}\!+\!j_{2}, &\!\!
\mid\! j_{3}\!-\!j_{4} \!\mid \leq J_{34} \leq j_{3}\!+\!j_{4} ,&\!\!
\mid\! j_{1}\!-\!j_{3} \!\mid \leq J_{13} \leq j_{1}\!+\!j_{3} , \vspace{0.1cm}  \\
\mid\! j_{2}\!-\!j_{4} \!\mid \leq J_{24} \leq j_{2}\!+\!j_{4} ,&\!\!
\mid\! J_{12}\!-\!J_{34} \!\mid \leq J \leq J_{12}\!+\!J_{34}  , &\!\!
\mid\! J_{13}\!-\!J_{24} \!\mid \leq J \leq J_{13}\!+\!J_{24} 
\end{array}
\end{equation}
are satisfied.

Tensor product of spinor spherical harmonics and/or spherical harmonics are expressed by
\begin{equation}
\begin{array}{l} 
\left[\mathcal{Z}_{\!s_{1}L_{1}}^{(J_{1})}(\!\Omega) \!\! \times \!\!
\mathcal{Z}_{\!s_{2}L_{2}}^{(J_{2})}(\!\Omega)\right]_{M}^{(J)}
\displaystyle \!\!\!\! \equiv \!\!\!\!
\sum_{(M_{1} M_{2})}\!\!\! \langle J_{1} M_{1} J_{2} M_{2} \!\!\mid \!\! JM \rangle
\mathcal{Z}_{s_{1}\!L_{1}\!M_{1}}^{(J_{1})}\!(\Omega)
\mathcal{Z}_{s_{2}\!L_{2}\!M_{2}}^{(J_{2})}\!(\Omega)
\vspace{0.3cm} \\ 
\displaystyle \!\!\!\!\!=\!\!\sum_{sL} \langle L_{1} 0 L_{2} 0 \! \mid \! L 0 \rangle \,
U\!\! \left( \begin{array}{ccc} \!\!\! s_{1} & \!\!\! L_{1} & \!\!\!J_{1} \\  \!\!\! s_{2} & \!\!\! L_{2} & \!\!\! J_{2} 
\\ \!\!\! s & \!\!\! L & \!\!\! J \end{array} \!\!\!\! \right)\!\!
\left[ \left[\sigma^{(s_{1})}\!\!\! \times \! \sigma^{(s_{2})} \right]^{(s)} \!\!\!\!\!\! \times \!
\left[C^{(\!L_{1})\!}(\Omega)\!\!\times\!\! C^{(\!L_{2}\!)}(\Omega)\right]^{\!(L)} \!\right]^{\!(J)}
\vspace{0.3cm} \\ \hspace{4.3cm}
 = \displaystyle  \sum_{sL} 
D(\lambda_{1} \! \lambda_{2} ; \lambda )  \mathcal{Z}_{sLM}^{(J)}(\Omega)
\end{array}
\end{equation}
The indices $\lambda_{1}$,$\lambda_{2}$, and $\lambda$ are respectively combined indices of
$\lambda_{i} =(s_{i} L_{i} J_{i} M_{i}) (i=1,2) $ and $\lambda=(s L J M)$.
Using the following relations
\begin{equation}
\begin{array}{l}
\left[ \sigma^{(0)} \times \sigma^{(s)} \right]^{(s)}_{\mu}\!=
\left[ \sigma^{(s)} \times \sigma^{(0)} \right]^{(s)}_{\mu}\!=\sigma^{(s)}_{\mu}, \hspace{0.4cm}
\left[ \sigma^{(1)} \times \sigma^{(1)} \right]^{(2)}_{\mu}\!=\vect{0} \vspace{0.3cm} \\
\left[ \sigma^{(1)} \times \sigma^{(1)} \right]^{(1)}_{\mu}\!=-\sqrt{2}\sigma^{(1)}_{\mu} \hspace{0.3cm}
\left[ \sigma^{(1)} \times \sigma^{(1)} \right]^{(0)}_{0}\!=-\sqrt{3} \sigma^{(0)}_{0}
\end{array}
\end{equation}
and 
\begin{equation}
\left[ C^{(L_{1})}(\Omega) \times C^{(L_{2})}(\Omega)\right]^{(L)}_{M}=
\langle L_{1} 0 L_{2} 0 \mid L 0 \rangle C^{(L)}_{M}(\Omega)
\end{equation}
and eq.(\ref{def_v}), the vertex coefficients $D(\lambda_{1} \lambda_{2} ; \lambda)$ are obtained for 
$s_{1}=s_{2}=s=1$ as
\begin{equation}
D(\lambda_{1} \lambda_{2} ;\lambda) \!= \! -\sqrt{6 (2L\!+\!1)(2J_{1}\!+\!1)(2J_{2}\!+\!1)}
\langle L_{1} 0 L_{2} 0 \! \mid \!\! L 0 \rangle
\left\{ \begin{array}{ccc} \!\!1 & \!\!L_{1} & \!\!J_{1} \\ 
\!\! 1 & \!\! L_{2} & \!\! J_{2} \\ \!\! 1 & \!\! L & \!\! J \end{array} \!\!\!\right\}
\end{equation}
and for $s_{1}=s_{2}=1, s=0$
\begin{equation}
D(\lambda_{1} \lambda_{2}; \lambda) \!= \! (-1)^{J_{1}+L_{2}+L}\sqrt{(2J_{1}\!+\!1)(2J_{2}\!+\!1)}
\langle L_{1} 0 L_{2} 0 \! \mid \!\! L 0 \rangle
\left\{ \begin{array}{ccc} \!\!L_{1} & \!\!J_{1} & \!\! 1 \\ 
\!\! J_{2} & \!\! L_{2} & \!\! L \end{array} \!\!\!\right\}
\end{equation}
where $\left\{ \! \begin{array}{ccc} \!\!\! \circ &\!\!\! \circ & \!\!\! \circ \vspace{-0.15cm} \\ \!\!\!
\circ &\!\!\! \circ & \!\!\! \circ \vspace{-0.15cm} 
\end{array} \!\! \!\right\}$  
stands for Wigner 6-j symbol.
The Clebsch-Gordan coefficient $\langle L_{1} 0 L_{2} 0 \mid L 0 \rangle \neq 0$ only the case of
$L_{1}+L_{2}-L={\rm even}$, indicating parity conservation.
Because of the symmetry property (3) of 9-j symbol, the vertex coefficients have the following
phase relation,
\begin{equation}
D(\lambda_{2} \lambda_{1};\lambda) =(-1)^{s_{1}+s_{2}+s+L_{1}+L_{2}+L+J_{1}+J_{2}+J}
D(\lambda_{1} \lambda_{2};\lambda) .
\end{equation} 
In the case of $s_{i} =0 , s_{3-i}=s=1 (i=1,2)$
\begin{equation}
D(\lambda_{1},\lambda_{2} ;\lambda_{3})=(-1)^{J_{i+1}+L_{i+1}+1}\sqrt{(2J_{i+1}+1)(2L_{i+1}+1)}
\langle L_{1} 0 L_{2} 0 \! \mid \!\! L 0 \rangle
\end{equation}
%
%     Appendix B 
\setcounter{equation}{0}
\setcounter{section}{2}
\begin{center} {\bf  \large Appendix B} \end{center}
\vspace{0.3cm}

The Dirac equation in free space is given for spherical symmetric potential
\begin{equation}
i\hbar \frac{\partial}{\partial t} \Psi(x)\!=\!\left( c_{0}\vect{\alpha}\cdot \vect{p} +U(r)\vect{1}_{4}
 +\beta mc^{2}\right) \Psi(x)\!=\!H\Psi(x)
\hspace{0.4cm} {\rm with} \hspace{0.2cm} \vect{p}\!=\!-i\hbar_{\ast}\vect{\nabla}
\end{equation}
Dirac field $\Psi(x \sigma)$  is expanded in terms of spinor spherical harmonics
$\mathcal{Y}_{\kappa \mu}(\theta \phi \sigma)$ based on $j$-$j$ cooupling scheme,\cite{He00} 
\begin{equation}
\Psi(x)=\!\!
\left( \begin{array}{c}  \psi^{u}(t\,\vect{r}\sigma) \vspace{0.2cm}  \\ \psi^{\ell} (t \, \vect{r} \sigma)
\end{array} \right) 
=\left( \begin{array}{l} 
\sum_{\kappa \mu} g^{(u)}_{\kappa \mu}(t \, r)\mathcal{Y}_{\kappa \mu}(\theta \phi \sigma) \, ,
 \vspace{0.2cm} \\ 
  \sum_{\kappa \mu} g^{(\ell)}_{\kappa \mu}(t \, r)\mathcal{Y}_{\tilde{\kappa} \mu}(\theta \phi \sigma) 
 \end{array} \right) \, ,
 \end{equation}
where
\begin{equation}
\mathcal{Y}_{\kappa \mu}(\theta \phi \sigma)=\sum_{\sigma = \pm \frac{1}{2} } 
\langle \ell_{\kappa} m \frac{1}{2} \sigma \mid j_{\kappa} \mu \rangle
(i)^{\ell_{\kappa}} Y_{\ell_{\kappa}m}(\theta \phi) \chi _{\sigma}^{(\frac{1}{2})}
\end{equation}
where quantum number $\kappa$ takes on non-zero integers relating to orbital angular momentum
 $\ell_{\kappa}$ and total angular momentum $j_{\kappa}$ in units of $\hbar_{\ast}$ as follows\cite{Ro37}
 \begin{equation}
 \mid \!\! \kappa \!\! \mid=k = j_{\kappa}+\frac{1}{2},  \hspace{0.3cm} \ell_{\kappa} \pm \frac{1}{2}
 \hspace{0.2cm}  {\rm for } \hspace{0.2cm} \kappa=\pm k \, ,
 \end{equation}
 and $\tilde{\kappa}=-\kappa$.
 The quantum number $\kappa$ is eigen-value of the operator 
 \begin{equation}
 K=\beta (\vect{\sigma} \cdot \vect{\ell} + \vect{1})
 \end{equation}
 and $k$ takes on natural number.
 It is proven that $[K.H]=0$ leading an eigen-value equation
 \begin{equation}
 H\Psi_{\kappa}^{(\nu)}\!=
 \! \left(\!\! \begin{array}{cc}  (U(r)+ mc_{0}^{2})\vect{1}    & c_{0}\vect{\sigma} \cdot \vect{p} 
 \vspace{0.2cm} \\   c_{0}\vect{\sigma} \cdot \vect{p}  & \hspace{-0.2cm} (U(r) - mc_{0}^{2})\vect{1} 
 \end{array}\!\! \right) \!
 \left( \!\! \begin{array}{l} 
g^{(u)}_{\kappa \mu}(r)\mathcal{Y}_{\kappa \mu}(\theta \phi \sigma) \, ,
 \vspace{0.2cm} \\ 
 g^{(\ell)}_{\kappa \mu}(r)\mathcal{Y}_{\tilde{\kappa} \mu}(\theta \phi \sigma) 
 \!\! \end{array} \right)\!=\!E_{\nu}\Psi_{\kappa}^{(\nu)}
  \end{equation}

We define a pseudo scalar operator 
\begin{equation}
\sigma_{r}= \vect{\sigma}\cdot \vect{n}_{r} \!=\!-\!\sqrt{3}\mathcal{Z}^{(0)}_{110}\!(\theta \phi)
, \hspace{0.5cm} 
\vect{n}_{r} = (\sin \theta \cos \phi , \sin \theta \sin \phi , \cos \theta ) ,
\end{equation}
having a property of $\sigma_{r}^{2} = \vect{1} $ and altering the parity without changing $j$
as
\begin{equation}
\sigma_{r} \mathcal{Y}_{\kappa \mu} =-i S_{\kappa} \mathcal{Y}_{\tilde{\kappa} \mu} .
\end{equation}
and we obtain an identity
\begin{equation}
\vect{\sigma}\!\cdot\!\vect{p}\!=\! \frac{\sigma_{r}}{r}
(\vect{\sigma}\!\cdot\! \vect{r})(\vect{\sigma}\!\cdot\!\vect{p}) 
\!=\!\frac{\sigma_{r}}{r} (\vect{r}\!\cdot\!\vect{p} \!+\! i\vect{\sigma}\!\cdot (\vect{r}\!\times\! \vect{p}))
\!=\! i\hbar_{\ast}\sigma_{r}\left(\!\!-\vect{1}\frac{\partial}{\partial r}+\vect{\sigma}\cdot \vect{\ell}\!\right) 
\end{equation}
We redefine radial wave function for convenience,
\begin{equation}
\left( G_{\kappa \mu}^{(u)}(r) , G_{\kappa \mu}^{(\ell)}(r) \right)
=\left( r g_{\kappa \mu}^{(u)}(r) , r g_{\kappa \mu}^{(\ell)}(r) \right)
\end{equation}
Dirac equation is expressed 
\begin{equation}
\frac{d}{dr}\!\left(\!\! \begin{array}{c} G_{\kappa \mu}^{(u)}(r) \vspace{0.2cm} \\ G_{\kappa \mu}^{(\ell)}(r)
\end{array} \!\!\!\! \right)\!=\!\left( \!\!\begin{array}{cc} \!\!\!
\displaystyle -\frac{\kappa}{r} &\displaystyle \hspace{-1.2cm} -\frac{1}{\hbar_{\ast}}(U(r)-E_{\nu} -mc_{0}^{2})
\vspace{0.2cm} \\ \displaystyle
\frac{1}{\hbar_{\ast}}(U(r)-E_{\nu} +mc_{0}^{2})& \displaystyle \frac{\kappa}{r} \end{array}  \!\!\!\! \right)\!\!
 \left(\!\! \begin{array}{c} G_{\kappa \mu}^{(u)}(r) \vspace{0.2cm} \\ G_{\kappa \mu}^{(\ell)}(r) 
\end{array}\!\!\!\! \right) 
\end{equation}
\vspace{0.3cm}
\begin{center} {\bf Aknowredgement}\end{center}\vspace{0.1cm}

The author would like to express hearty thanks to Professor Tetsuo Hyodo for fruitful discussions 
and for letting him use facilities in the laboratory. He apologize to Professor Tomoyuki Maruyama,
because the author has left halfway finished the collaborated work making the subject on
"Origin of Strong Toroidal Magnetic Field in Magnetar". In that work, the spin current of electrons
was treated with assumption of a given proton distribution\cite{OM20}. 
The present work intends to treat global structure of magnetic
field including toroidal field without any assumption on the distribution of protons and to complete that work.

\end{document}